
\input harvmac.tex
\noblackbox

\let\absrm\tenrm

\def\onebyhbar{{\textstyle{1\over\hbar}}}
\def\ibyhbar{{\textstyle{i\over\hbar}}}
\def\to{\rightarrow}

\def\CO{{\cal O}}

\def\ANP#1{{\sl Ann. Phys.}        {\bf  #1}}
\def\CMP#1{{\sl Comm. Math. Phys.} {\bf  #1}}
\def\IJTP#1{{\sl Int. J. Th. Phys.}{\bf  #1}}

\def\NCX#1{{\sl Nuovo Cimento    } {\bf  #1}}
\def\NPB#1{{\sl Nucl. Phys.}       {\bf B#1}}
\def\PHY#1{{\sl Physics}           {\bf  #1}}

\def\PRX#1{{\sl Phys. Rev.}        {\bf  #1}}

\def\PRL#1{{\sl Phys. Rev. Lett.}  {\bf  #1}}
\def\RMP#1{{\sl Rev. Mod. Phys.}   {\bf  #1}}

\line{\hfill IISc-CTS-5/93}
\line{\hfill hep-th/9309077}
\bigskip

\centerline{\bf ANOTHER LOOK AT BELL'S INEQUALITIES}
\centerline{\bf AND QUANTUM MECHANICS\footnote{$^\dagger$}{%
%
Contributed to the XVI International Symposium on Lepton-Photon
Interactions, Cornell University, August 10-15, 1993.}}

\bigskip
\centerline{APOORVA PATEL\footnote{$^*$}{E-mail: adpatel@cts.iisc.ernet.in}}
\medskip
\vbox{\it\baselineskip=12pt
\centerline{CTS and SERC, Indian Institute of Science}
\centerline{Bangalore-560012, India}}

\bigskip
\centerline{ABSTRACT}
\medskip
\vbox{\absrm\baselineskip=12pt\narrower\noindent
Feynman's path integrals provide a hidden variable description of
quantum mechanics (and quantum field theories). The expectation values
defined through path integrals obey Bell's inequalities in Euclidean
time, but not in Minkowski time. This observation allows us to pinpoint
the origin of violation of Bell's inequalities in quantum mechanics.}

\bigskip
\vbox{\absrm\narrower\noindent PACS numbers: 03.65.Bz, 03.65.Db}
\bigskip

It is common, and often more convenient, to study quantum mechanics
using the Schr\"odinger/Dirac equations and the Heisenberg operator
algebra. Feynman's path integral formulation of quantum mechanics
\ref\pathint{R. Feynman, \RMP{20} (1948) 367
    \semi    R. Feynman and A. Hibbs, {\it Quantum Mechanics
             and Path Integrals} (McGraw-Hill, New York, 1965).},
nonetheless, is an alternative approach from which all the known
results of quantum mechanics can be derived. In fact, quantum field
theories are studied, more often than not, using the path integral
formulation. For simplicity, let us consider quantum mechanics in
one space dimension. The total amplitude (or wavefunction) for the
system to be in the state $\psi(x_f,T)$ at time $t=T$, given an
initial state $\psi(x,0)$, is defined in terms of the transition
kernel $K(x_f,T;x_i,0)$:
\eqn\wavefn{
\psi(x_f,T) ~=~ \int_{-\infty}^\infty K(x_f,T;x_i,0)
              ~ \psi(x_i,0) ~dx_i ~~,
}
\eqn\kernel{
K(x_f,T;x_i,0) ~=~ \int_0^T [Dx(t)] ~\exp [\ibyhbar S(x(t))] ~~.
}
The integration measure $[Dx(t)]$ can be defined precisely by
discretising the time interval:
\eqn\dxmeas{
[Dx(t)] ~\propto~ \lim_{N\to\infty} \int_{-\infty}^\infty
         \prod_{j=1}^{N-1} dx_j ~~,~~ x_j ~\equiv~ x(t=jT/N) ~~.
}
Transition matrix elements and other expectation values are defined
with the straighforward prescription:
\eqn\expectval{
\vev{\CO(x(t))} ~\equiv~ \vev{\psi(x_f,T) | \CO(x,t) | \psi(x_i,0}
                     ~=~ \int [Dx(t)] ~\rho_M (x(t))~ \CO ~~,
}
\eqn\minkmeas{
\rho_M (x(t)) ~=~ \exp [\ibyhbar S(x(t))] /
                  \int [Dx(t)] ~\exp [\ibyhbar S(x(t))] ~~.
}

These definitions provide a hidden variable description of quantum
mechanics. Indeed, $x_j~(j=1,...,N-1)$ are the hidden variables
which are integrated over. Note that $\{x_j\}$ corresponding to
one particle are totally independent of those for another; all the
correlations are built into specification of the initial state
coordinates, $x_0$. There is no need to worry about ordering of
various factors, because there are no non-commuting operators in
this Lagrangian description, only complex numbers\foot{%
Fermions can be dealt with along similar lines using Grassmann
variables, but I do not consider them here.}.
The functional integration measure $[Dx(t)]$ represents a sum over
all paths connecting the fixed initial and final states and is
inherently non-local\foot{%
Quantum fluctuations can be suppressed and locality can be restored
by taking the formal limit, $\hbar\to0$.}.
$\exp[\ibyhbar S(x(t))]$ is the statistical weight of the path $x(t)$
which depends on the interaction amongst the particles in the system.
It is well-known that the typical paths contributing to the functional
integral are highly irregular and non-differentiable. The individual
paths characterised by $\{x_j\}$ follow local history/dynamics\foot{%
This is true in quantum mechanics only for non-interacting particles,
since long range inter-particle interaction potentials would produce
instantaneous action-at-a-distance. Locality is much more general,
however, in quantum field theories describable in terms of local
Lagrangian densities.},
but they do not obey constraints of causality. The virtual
intermediate states, for example, can be ``off-shell'' with no
relation between their energies and momenta, and they can propagate
at a speed faster than that of light. They can even propagate
backwards in time which is interpreted as pair creation and
annihilation in relativistic field theories. The total amplitude
(i.e. the sum over all paths), however, obeys all the constraints
of causality and conservation laws.

The measurement process here corresponds to restricting the sum over
all possible paths to only those paths which are consistent with the
measured observable having a specific value in the final state. This
is obviously a contextual process, once all the hidden variables are
integrated out; if a second measurement were to be carried out on the
system, the sum over all paths would be restricted to paths which
are consistent with the results of both the first and the second
measurements---the paths which would be consistent with the second
measurement but not with the first have been discarded by the act
of the first measurement. Thus the measured value of an observable
may depend on the results of other measurements carried out prior
to it on the same system, when these other measurements correspond
to commuting but correlated observables. Note that the order in which
the restriction of paths is carried out is immaterial, so there is no
ambiguity regarding the final state of the system when measurements
are carried out at space-like separations.

All put together, path integrals describe a contextual and non-local
hidden variable theory\foot{%
Bell's theorems ensure that any hidden variable theory describing
quantum mechanics has to have such peculiarities
\ref\mermin{A recent review on the implications of Bell's theorems can
            be found in: N. D. Mermin, \RMP{} (1993) to appear.}.}.
It is similar in many aspects to the de Broglie-Bohm theory
\ref\bohm{D. Bohm, \PRX{85} (1952) 166;180.},
but with the clear advantage that it is describable in a local language
and that it is much more amenable to detailed calculations\foot{%
In particular, we know how to generalise from non-relativitistic to
relativistic case, and from quantum mechanics to quantum field theory.}.
The crucial feature in this description is that the integration weight,
$\rho_M (x(t))$, is a complex number in general. Therefore, although it
is bounded, it cannot be interpreted as a probability density. It is
this fact which allows the path integral description to bypass Bell's
inequalities
\ref\bellineq{J. Bell, \PHY{1} (1964) 195
    \semi     J. Bell, {\it Speakable and Unspeakable in Quantum
              Mechanics} (Cambridge University Press, Cambridge, 1987).}
and give a true definition of quantum mechanics\foot{%
For example, there can be cancellation between different paths giving
rise to interference effects.}.
Feynman actually showed how one can reconcile the EPR paradigm
\ref\eprparadigm{A. Einstein, B. Podolsky and N. Rosen,
                 \PRX{47} (1935) 777
    \semi        A. Einstein, in {\it Albert Einstein, Philosopher
                 Scientist}, Ed. P. Schilp, Library of Living
                 Philosophers (Evanston, Illinois, 1949).}
with quantum mechanics using hidden variables, provided that
probabilities are allowed to become negative
\ref\negprob{R. Feynman, \IJTP{21} (1982) 467
    \semi    R. Feynman, ``{\it Negative probability}'' in
             {\it Quantum Implications: Essays in honour of David Bohm},
             Ed. B. Hiley and F. David Peat, p.235
             (Routledge \& Kegan Paul, London, 1987).}.
Wigner functions
\ref\wigner{E. Wigner, \PRX{40} (1932) 749.}
describing the density matrix distribution in the phase space are
well-known examples of this type:
\eqn\wignerfun{
W(x,p) ~=~ \int ~ \psi^* (x+\half y) ~\exp (\ibyhbar py)
                ~ \psi   (x-\half y) ~~.
}
They are real and obey all the usual manipulations of probability
theory except that they are not always positive\foot{%
My use of ``positive'' in this article includes zero, i.e. it really
means ``non-negative''.}
everywhere in the phase space. Since the expectation value for any
physical observable is just
\eqn\wignerexpval{
\vev{\CO} ~=~ \int dx~dp ~ W(x,p) ~ O(x,p) ~~,
}
where $O(x,p)$ is the (Hermitian) operator weight corresponding to the
observable $\CO$, it is necessary that for situations violating Bell's
inequalities the Wigner function be negative somewhere in the phase space.

Now let us use the familiar trick of rotating to Euclidean (imaginary)
time, $\tau=it$. This rotation converts the quantum theory to the
language of statistical mechanics. The integration weight,
\eqn\euclmeas{
\rho_E (x(\tau)) ~=~ \exp [-\onebyhbar S(x(\tau))] /
                     \int [Dx(\tau)] ~\exp [-\onebyhbar S(x(\tau))] ~~,
}
is now real and positive, and can be interpreted as a probability
density. This formal property has been exploited before, for
vector-like gauge theories such as QCD, to derive rigorous
inequalities among correlation functions and particle masses
\ref\weingarten{D. Weingarten, \PRL{51} (1983) 1830.}
\ref\vafawitten{C. Vafa and E. Witten, \NPB{234} (1984) 173\semi
                E. Witten, \PRL{51} (1983) 2351.}.
The point I want to emphasise in this paper is that a positive
integration weight must obey Bell's inequalities\foot{%
I use the term ``Bell's inequalities'' in this article to generically
denote the type of inequalities proved by Bell as well as by others.}
Something has been lost in the analytic continuation from Minkowski
to Euclidean time and the question is to identify what it is.

Let us first note that the above mentioned analytic continuation
is routinely employed in quantum field theories in dealing with
divergent loop integrals and renormalisation. As a matter of fact,
there are strong theorems governing such an analytic continuation,
e.g. the Wightman axioms
\ref\wightman{R. Streater and A. Wightman, {\it PCT, spin and statistics,
              and all that} (Benjamin/ Cummings, New York, 1964).}
and the Osterwalder-Schrader positivity of the transfer matrix
\ref\osterwalder{K. Osterwalder and R. Schrader, \CMP{31} (1973) 83;
                 \CMP{42} (1975) 281
     \semi       K. Osterwalder and E. Seiler, \ANP{110} (1978) 440.}.
In particular, in the complex energy plane, these theorems rely on a
sufficiently fast decrease of the amplitudes at infinity and on there
being no singularities in the region covered by the rotation\foot{%
The poles corresponding to unstable states or resonances lie on the
second Riemann sheet.}.
The integration contours can then be freely rotated without affecting
the value of the integrals.

Let us also note that statistical mechanics has many features in
common with quantum theory. The density matrix description allows
probabilistic interpretation of superposed and mixed states. The
presence of the heatbath\foot{%
$\hbar\ne0$ corresponds to non-zero temperature.}
gives rise to vacuum fluctuations and unconstrained behaviour of the
virtual states. Non-zero commutators (Poisson brackets) leading to
the uncertainty principle, e.g. $[p,x]\ne0$, exist in the Hamiltonian
description of the theory. Non-zero tunnelling amplitudes exist in the
Euclidean theory, and so do ``grotesque'' states with infinitesimal
probabilities\foot{%
For instance, there is a non-zero probability for all the molecules
of air in a room to collect themselves in one corner, but it would
be silly to wait for such a thing to happen spontaneously.}.
These shared features cannot be responsible for the violation of
Bell's inequalities in quantum mechanics.

What the Euclidean theory lacks is the concept of unitarity. The
Minkowski space transition kernel, $K_M (x_f,T;x_i,0)$, corresponds
to a unitary transformation---the familiar $S-$matrix. It is easy
to see that a unitary matrix with only real and positive matrix
elements has to be the identity matrix (or its row-wise permutation
corresponding to a shuffling of the states). The Euclidean space
transition kernel is less restricted---it does not preserve the
norm, though it maintains orthogonality of the states---and can be
represented by a diagonal positive definite matrix. The loss of
normalisation is not critical, since it can be taken care of
following the LSZ prescription
\ref\lszred{H. Lehmann, K. Symanzik and W. Zimmermann,
            \NCX{1} (1955) 1425.}.
The information about the relative phases of the states, however,
is lost. In fact this is the crucial quantum mechanical feature
which makes the requirements of unitarity and a real positive
integration weight mutually incompatible\foot{%
Feynman's choice of negative probabities corresponds to an
orthogonal transformation instead of a unitary one. In special
circumstances, constraints of orthogonality are sufficient to
show that Bell's inequalities are violated in quantum mechanics.}.
One can pick situations where the information contained in the
relative phases cannot be made arbitrary small, because there are
constraints on the complex amplitudes following, for example, from
analyticity and dispersion relations. In such cases a statistical
mechanics description cannot provide an arbitrarily close
approximation to quantum mechanics. Bell's inequalities are
explicit examples of this type\foot{%
Normally Bell's inequalities are not looked upon in this manner.
They are instead specified by certain correlations in the initial
state of a non-interacting system. In the real world, say in an
$S-$matrix example, the correlations are not present in the initial
state, but are generated dynamically by interaction phases.}.

Let us look at the problem again from a slightly different angle.
The Euclidean correlation functions defined along the imaginary
time axis are real, and a mere analytic continuation of them
cannot produce non-trivial complex phase shift factors that are an
an essential part of Minkowski scattering amplitudes\foot{%
It is still possible to extract phase shifts from Euclidean results
using indirect methods, e.g. finite volume shifts in energy levels
of a multi-particle system
\ref\euclphshift{M. L\"uscher, \CMP{105} (1986) 153;
                 \NPB{354} (1991) 531.}.}.
{\it A priori} one does not know whether in analytical continuation
of the Euclidean results the Euclidean time axis should be rotated
by $+90^\circ$ or $-90^\circ$ to reach the Minkowski time axis.
This ambiguity automatically gets resolved for internal loop
variables of Feynman diagrams, just due to the necessity of not
crossing any singularities while rotating the integration contours.
For the variables corresponding to the external legs (incoming
and outgoing states), however, the choice must be enforced as an
additional condition; it amounts to the difference between choosing
advanced or retarded propagators\foot{%
Given Eucliidean time results, one can construct either advanced
or retarded (or a linear combination thereof) amplitudes. The
choice of retarded amplitudes only, though physically motivated
by principle of causality, is absent in the Euclidean theory and
has to be used as an additional input.}.

More explicitly, the principle of causality is embedded in the use
of $i\epsilon-$prescription for retarded propagators. For example,
the naive analytic continuation of the Euclidean scalar boson
propagator, $(k^2+m^2)^{-1}$, to $(k^2-m^2)^{-1}$ is incorrect.
The correct physical prescription is:
\eqn\propcont{\eqalign{
{1 \over k^2+m^2} ~\to&~ \lim_{\epsilon\to0}
                         {1 \over k^2-m^2+i\epsilon} \cr
                    ~=&~ {\rm P} \big( {1 \over k^2-m^2} \big)
                         - i\pi\delta(k^2-m^2) ~~. \cr}
}
The non-unitary Euclidean theory fixes the off-shell amplitude
(the principle value part) completely, but it is necessary to add
an on-shell contribution (exemplified by the $\delta-$function at
the pole) to comply with unitarity. This subtlety is unimportant
in many cases: location of poles and branch cuts, matrix elements
with external legs amputated according to the LSZ prescription,
scattering amplitudes at threshold which are real etc. can be
determined without recourse to the $i\epsilon-$prescription. On the
other hand, $S-$matrix phase shifts, discontinuities across branch
cuts etc. are not directly accessible in the Euclidean framework.

The conventional set up illustrating Bell's inequalities is the case
where there is just free propagation of particles after initial
creation of the state. Let us first look at the original EPR example
\eprparadigm---a system of two identical non-relativistic particles
freely propagating in one space dimension, where the initial state
of the two particles is perfectly correlated in space (hence
anti-correlated in momentum) with the constraint
\eqn\corrdelta{
\delta (x_i-y_i) ~=~ \int {dp \over 2\pi\hbar}
                   ~ \exp \big( \ibyhbar p(x_i-y_i) \big) ~~.
}
For a single free particle of mass $m$, the propagation kernel for
Minkowski time $T$ is
\eqn\propkernM{
K_M (x_f,T;x_i,0) ~=~ \sqrt{m \over {2\pi i\hbar T}}
  ~ \exp \big( i{m(x_f-x_i)^2 \over {2\hbar T}} \big) ~~,
}
while for Euclidean time $T$ it is
\eqn\propkernE{
K_E (x_f,T;x_i,0) ~=~ \sqrt{m \over {2\pi\hbar T}}
  ~ \exp \big( -{m(x_f-x_i)^2 \over {2\hbar T}} \big) ~~.
}
Thus the final state wavefunction for the two particle system is
\eqn\wavefnM{
\psi_M (x_f,y_f,T) ~=~ \int {dp \over 2\pi\hbar}
                       \exp \big( \ibyhbar p(x_f-y_f) \big)
                       \exp \big( {ip^2 T \over \hbar m} \big)
}
in Minkwoski time, and
\eqn\wavefnE{
\psi_E (x_f,y_f,T) ~=~ \int {dp \over 2\pi\hbar}
                       \exp \big( \ibyhbar p(x_f-y_f) \big)
                       \exp \big( -{p^2 T \over \hbar m} \big)
}
in Euclidean time. In both cases the theory maintains its
contextual character; the non-locality of the kernel ensures
that if one particle is detected with mromentum $p_f$,
the other is bound to be found with momentum $-p_f$.

The relative probability of observing various values of momentum
is different in the two cases, however, which is just due to the
difference in normalisation of states. This difference is not small
as can be seen by appealing to the uncertainty relation: In order
to detect the two particles distinctly, their separation has to be
much larger than their de Broglie wavelengths,
\eqn\mindist{
|x_f-y_f| ~\sim~ p_f T/m ~>>~ \hbar/p_f ~~,
}
implying that $p^2 T >> \hbar m$ in the exponent. In this particular
example, the exponent can be removed by just following the LSZ
prescription, and there is no conflict with any inequality.

The reason behind no conflict with any inequality in the above
example is that the $\delta-$function correlation is positive.
A more general case will have initial state correlations such that
the Wigner function becomes negative somewhere in the phase space\foot{%
A density matrix with negative entries is perfectly alright for
quantum systems, though it would be anathema in statistical physics.}
\ref\negwigner{See for example: J. Bell, in {\it New Techniques and
               Ideas in Quatum Measurement Theory} (1986); reproduced
               as Chapter 21 in {\it Speakable and Unspeakable in
               Quantum Mechanics}, Ref. \bellineq.}.
The density matrix evolves linearly in time according to:
\eqn\denmatevol{\eqalign{
{d W_M \over dt}    ~=~ -\ibyhbar   [H,W_M] ~~\Longrightarrow~~&
  W_M(t)    ~=~ \exp(-\ibyhbar   Ht)    W_M(0) ~~, \cr
{d W_E \over d\tau} ~=~ -\onebyhbar [H,W_E] ~~\Longrightarrow~~&
  W_E(\tau) ~=~ \exp(-\onebyhbar H\tau) W_E(0) ~~, \cr}
}
where $H$ is the Hamiltonian for the system.
In case of a positive time evolution kernel, interference effects can
only annihilate the negative density matrix regions in the phase space;
a limit is reached when the Wigner function becomes positive everywhere
and thereafter no regions of negative Wigner function can be
regenerated\foot{%
Though we are talking about the density matrix here, the situation is
quite analogous to the evolution of the wavefunction in Euclidean time:
It is common knowledge that the ground state wavefunction in quantum
mechanics has no nodes. Euclidean time evolution concentrates the
wavefunction towards its ground state component, cutting down the
excited state contributions responsible for nodes in the wavefunction.}.
Explicitly
\eqn\wignertime{
f(\tau) ~=~ \int dx~dp ~ |W(x,p;\tau)| ~/~ \int dx~dp ~ W(x,p;\tau) ~~,~~
  f(\tau) \ge 1 ~~,~~ df(\tau)/d\tau \le 0 ~~.
}
No such reduction of negative density matrix regions is expected in case
of a unitary time evolution; in fact the Minkowski time evolution of the
density matrix obeys the continuity equation \wigner:
\eqn\conteqn{
{\partial W(x,p) \over \partial t}
  + {p \over m} {\partial W(x,p) \over \partial x} ~=~ 0 ~~.
}
Thus, after factoring out the normalisation of states, the residual
correlations which typify Bell's inequalities, become substantially
different in the Euclidean case from the corresponding Minkowski values.

In brief, the Euclidean time evolution cannot dynamically create
correlations violating Bell's inequalities if such correlations are
absent in the initial state. Moreover, it wipes out such correlations
even when they are inserted by hand in the initial state.

Violation of Bell's inequalities in quantum mechanics is often
demonstrated using two non-relativistic spin-$\half$ particles
in a singlet state
\ref\spinsinglet{D. Bohm and Y. Aharanov, \PRX{108} (1957) 1070.}.
%
%
The Wigner function for this state has negative elements \negprob.
In this case, there are two ways of modifying the problem so that
Bell's inequalities are satisfied. The first is to rewrite the
spin operator in terms of the integration variables\foot{%
Spin after all is a property of the Poincar\`e group transformations.}.
(e.g. as the Pauli-Lubansk\`{\i} $4-$vector,
$m s^\mu = \half\epsilon^{\mu\nu\sigma\tau} p_\nu J_{\sigma\tau}$,
where $J_{\sigma\tau}$ are the Lorentz generators),
and then rotate from Minkowski to Euclidean time as described above\foot{%
Note that vector-like gauge theories have positive path integral
measure in Euclidean space.}.
An alternative is to replace the unitary spin-$SU(2)$ group by a
``real positive counterpart''.
%
%
Both these choices give up unitarity, though in different manners.
All this is, however, a subject for future work.

Finally, it is impertinent to ask why nature opted for unitarity or
the Minkowski metric. What this paper aims to demonstrate is that
it is the ubiquitous appearance of ``$i$'' together with a non-zero
value of ``$\hbar$'', so characteristic of quantum physics, that sets
it apart from classical physics: ``$i$'' causes violations of Bell's
inequalities, while ``$\hbar\ne0$'' is responsible for non-locality.
A lesson to be learnt is that care must be exercised in converting
results of Euclidean field theory (in particular attempts towards the
quantum theory of gravity) to physical amplitudes; the restrictions
following from unitarity have to be kept in mind.
In fact the imposition of unitarity constraints on Euclidean results
may provide a way to quantify the violations of Bell's inequalities.

\bigskip\noindent
{\bf Acknowledgements:}
I am grateful to Stephen Sharpe for arguments regarding final state
interactions in Euclidean field theory which led to this work.
I thank N. Mukunda for discussions.

\listrefs
\vfill\bye